\documentclass{iau_FM}
\usepackage{graphicx}

\title[FM 17.~~Advances in Stellar Physics from Asteroseismology] 
{Serendipitous Science from the K2 Mission}

\author[D. L. Buzasi {\it et al.}]   
{Derek L. Buzasi$^1$, Lindsey Carboneau$^1$, Carly Hessler$^1$, Andy Lezcano$^1$, \and Heather Preston$^{1,2}$}

\affiliation{$^1$Dept. of Chemistry \& Physics, Florida Gulf Coast University, \\10501 FGCU Blvd. South, Fort Myers, FL 33919 USA \\email: {\tt dbuzasi@fgcu.edu}\\[\affilskip] $^2$ Calusa Nature Center \& Planetarium, 3450 Ortiz Ave., Fort Myers, FL, 33905 USA}

\pubyear{2015}
\jname{Astronomy in Focus, Volume 1} 
\editors{Piero Benvenuti, ed.}
\begin{document}

\maketitle

\begin{abstract}
The K2 mission is a repurposed use of the Kepler spacecraft to perform high-precision photometry of selected fields in the ecliptic. We have developed an aperture photometry pipeline for K2 data which performs dynamic automated aperture mask selection, background estimation and subtraction, and positional decorrelation to minimize the effects of spacecraft pointing jitter. We also identify secondary targets in the K2 ``postage stamps'' and produce light curves for those targets as well. Pipeline results will be made available to the community. Here we describe our pipeline and the photometric precision we are capable of achieving with K2, and illustrate its utility with asteroseismic results from the serendipitous secondary targets.
\end{abstract}

\firstsection 
\section{Introduction}

The impressive rescue and repurposing of the K2 mission (\cite[Howell et al. (2014)]{Howell14}) has the promise of high-precision photometry 
for numerous ecliptic fields of view, with access to numbers and types of targets not envisioned in the original
Kepler mission. However, the operating mode for K2 differs significantly from that of the original Kepler mission,
so the data present significantly greater analysis complexities. This situation is exacerbated by the fact that
at present the K2 Guest Observer office does not plan to provide a vetted and standardized pipeline product, but 
instead simply access to the relatively raw observing frames. 

The lack of a pipeline and data product endorsed by the mission has led a number of teams to develop their own
pipelines 
(\cite[Aigrain et al. (2015)]{Aigrain15},
\cite[Armstrong et al. (2015)]{Armstrong15},
\cite[Lund et al. (2015)]{Lund15},
\cite[Vanderburg \& Johnson (2015)]{Van15}),
and the work presented here is broadly similar. However, since there is no ``ground truth'' 
for K2 data products, and different approaches to K2 data reduction are independent of one another, the existence of products
from multiple pipelines is complementary rather than competitive, and the differing approaches are likely to be more or less
suited to different kinds of targets. In addition, we are further motivated by the desire to perform serendipitous science using the background sources in each K2 frame, and by the desire to include undergraduate students in the development and analysis work. In some cases, we compare our results to those of \cite[Vanderburg \& Johnson (2015)]{Van15}, as this is currently the most widely-used pipeline.

\section{Pipeline}

Here we briefly describe the operation of the pipeline and its products for long-cadence K2 data. While operation and output shown are generally for the pipeline operating in batch mode, it can also be run in single-frame mode to obtain more refined results, particularly for targets displaying unusual photometric characteristics.

{\underline{\it Preprocessing \& Target Identification}}

Long-cadence K2 time series, like the Kepler time series, have 30-minute cadence, and the spacecraft returns ``postage stamps'' surrounding primary targets. These small images have varying sizes, typically approximately 25 pixels square in Campaign 0, but with decreasing sizes in following Campaigns. We begin by estimating and removing the background level in each image, using the median of the faintest 20\% of the pixels, and then construct a mean image from the time series as shown in Figure~1 for the Campaign 0 primary target EPIC~202066811. Due to the $\approx 6$ hourly spacecraft pointing corrections, stellar images show varying degrees of elongation depending primarily on the distance of the postage stamp from boresight.

\begin{figure}
\begin{center}
 \includegraphics[width=3.3in]{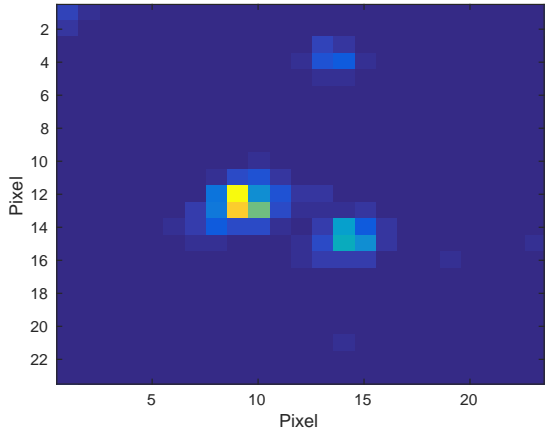} 
 \caption{The mean image from a K2 time series for EPIC 202066811 from Campaign 0. Multiple targets are clearly visible, as is the asymmetry typical
of K2 stellar images, due to the ~6 hourly pointing corrections of the spacecraft.}
   \label{fig1}
\end{center}
\end{figure}

Targets are identified in the mean image through a brute-force search for local maxima in the frame, avoiding edge pixels. This is a relatively inefficient process which is nonetheless
practical due to the small frame size. To avoid accidentally identifying background noise peaks as sources, as a default we require that
each peak be more than $0.5\sigma$ above its immediate neighbors, though this level can be manually modified in single-frame mode. The example shown in Figure~2 imposed a $1\sigma$ requirement, leading to 3 targets in the frame rather than the 5 which would result in default mode.

\begin{figure}
\begin{center}
 \includegraphics[width=3.3in]{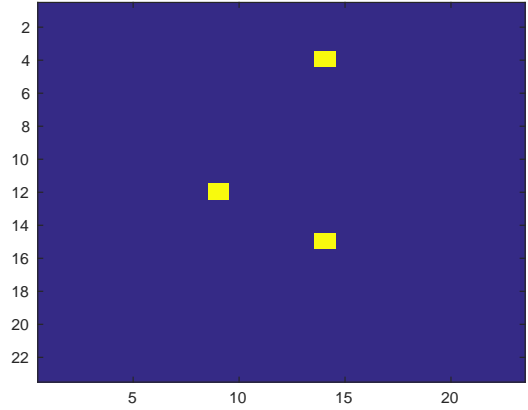} 
 \caption{Targets identified in the frame shown in Figure~1.}
   \label{fig2}
\end{center}
\end{figure}

{\underline{\it Mask Generation}}

We next generate a mask to use for aperture photometry. The initial mask for a target is simply the one-pixel local maximum previously identified as the target location, and a time series is calculated based on that pixel. The resulting time series is then twice-differenced to improve stationarity (\cite[Box \& Jenkins (1970)]{Box70}), and $\sigma$ is taken as a figure of merit. Additional mask are then generated algorithmically by adding the largest contiguous pixel. A new time series and figure of merit are then calculated, and the mask resulting in the time series with the lowest twice-differenced $\sigma$ is adopted. The process can be terminated early if adding an additional pixel would result in the addition to the mask of a pixel with a larger mean image value than any previously included in the mask. This latter constraint helps to avoid ``bridging'' between nearby targets. 

\begin{figure}
\begin{center}
 \includegraphics[width=3.3in]{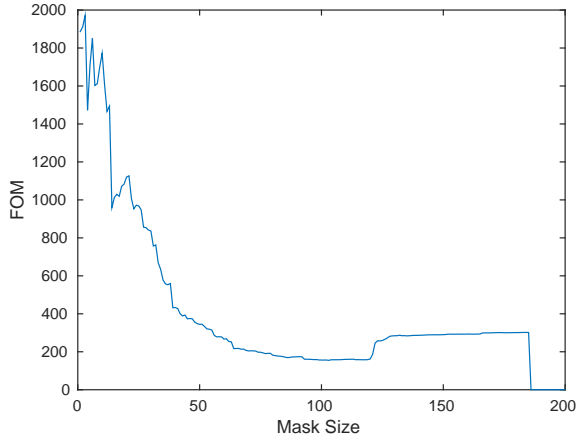} 
 \caption{The figure of merit used by the pipeline ($\sigma$ of the twice-differenced time series), shown as a function of mask size for a K2 Campaign 1 target. The best mask size in this case was 103 pixels, which is typical of targets in the uncrowded Campaign~1 frames; C0 mask sizes are generally smaller (see Figure~4).}
   \label{fig3}
\end{center}
\end{figure}

No attempt is made to ensure that masks for individual targets in a frame do not overlap. However, the mask construction procedure makes this uncommon and in practice it only occurs for truly overlapping targets. When masks do overlap, the number of overlapping pixels is noted and used as a proxy for crowding. Once the masks are constructed, simple aperture photometry is used to calculate fluxes along with $(x,y)$ centroid and background estimates. Figure~4 illustrates the resulting masks for the frame in Figure~1. 

\begin{figure}
\begin{center}
 \includegraphics[width=3.3in]{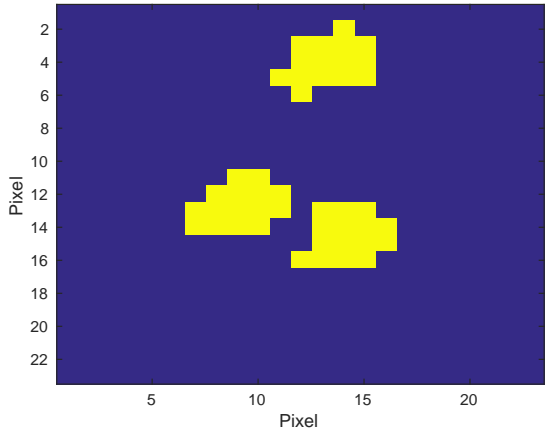} 
 \caption{Aperture photometry masks for the targets identified in Figure~2. Note that the derived mask sizes in this more-crowded field are significantly smaller ($\approx 20$ pixels) than those typical of the uncrowded C1 field seen in Figure~3.}
   \label{fig4}
\end{center}
\end{figure}

{\underline{\it Detrending}}

Detrending is normally required due to the presence of spacecraft pointing resets, along with additional correlated noise. We have the capability to detrend against $(x,y)$ centroid, background level, and FWHM, but normally only positional detrending is used. Points corresponding to spacecraft pointing resets are identified through examination of the differenced centroid time series and removed, and the entire time series is then detrended using a low-order (typically $n=3$) polynomial fit. The time series is then broken into overlapping segments of length $\Delta t \geq 1~\rm d$ (variations are possible depending on the behavior of the time series DFT), and further detrended using low-order ($n = 1-3$) polynomial fits, with the exact order a function of the number of points in each segment. Detrending is repeated in an iterative fashion until the improvement in successive time series second-differenced $\sigma$ fails to exceed 1\%; typically $3-10$ times. Figure~5 shows the typical improvement from raw to detrending time series, while Figure~6 illustrates the effect of detrending on the flux-centroid correlation.

\begin{figure}
\begin{center}
 \includegraphics[width=3.3in]{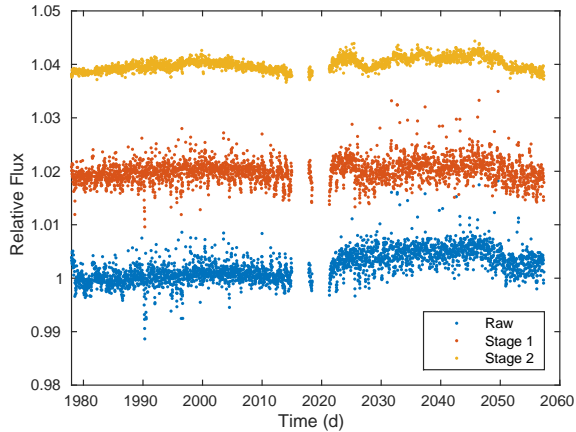} 
 \caption{Improvement in time series quality between raw, stage 1 detrending (entire time series), and stage 2 detrending (segmented time series). The overall offsets between the time series are added to aid visibility.}
   \label{fig5}
\end{center}
\end{figure}

\begin{figure}
\begin{center}
 \includegraphics[width=3.3in]{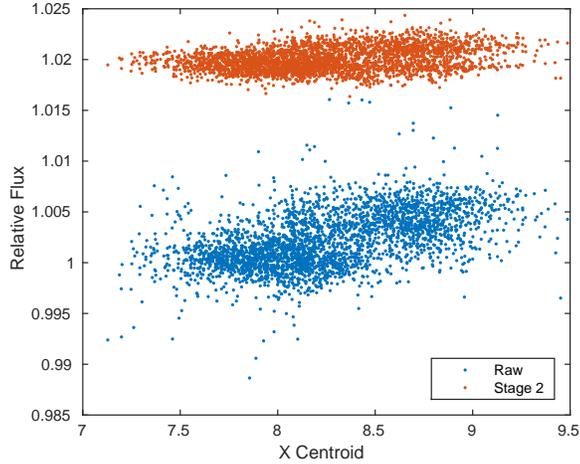} 
 \caption{Decorrelation effectively removes the relation between flux and centroid position which is present in the raw time series. The two data sets are arbitrarily offset to aid visualization.}
   \label{fig6}
\end{center}
\end{figure}

\section{Results}

The resulting time series have noise characteristics (particularly CDPP, \cite[Gilliland et al. (2011)]{Gilliland11}) which are similar to those shown by Vanderberg \& Johnson (2014), though typically our low-frequency noise is modestly better and our high-frequency noise is slightly worse. One aspect of this behavior, shown in Figure~7 for the $K_{mag} = 12.8$ Campaign 1 red giant star EPIC~201757842, is that the offsets between data before and after time series gaps which frequently occur in the Vanderberg \& Johnson (2014) pipeline are absent from ours. In addition, we tend to produce cleaner time series for fainter targets, but Vanderberg \& Johnson's results are generally superior for bright targets and those with larger-amplitude periodic variability.

\begin{figure}
\begin{center}
 \includegraphics[width=3.3in]{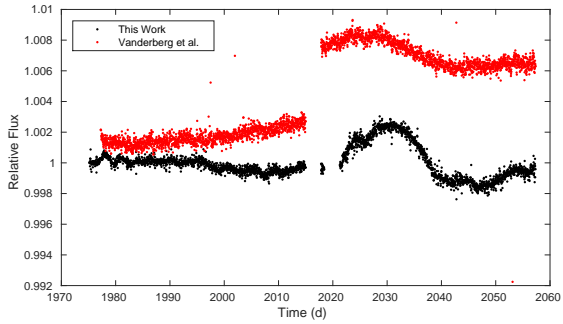} 
 \caption{A comparison of time series from our pipeline (black) and Vanderberg \& Johnson (2014, in red) for the same red giant target (EPIC 201757842). The power spectrum for this target is shown in Figure~8.}
   \label{fig7}
\end{center}
\end{figure}

Figure~8 illustrates the DFT for the time series shown in Figure~7, which should be compared with Figure~2 in Stello et al. (2015). A simple analysis broadly following the procedure in Stello et al. (2015) leads to $\nu_{max} = 148 \mu Hz$ and $\Delta \nu = 14 \mu Hz$, comparable (within errors) to the results in Stello et al.

\begin{figure}
\begin{center}
 \includegraphics[width=3.3in]{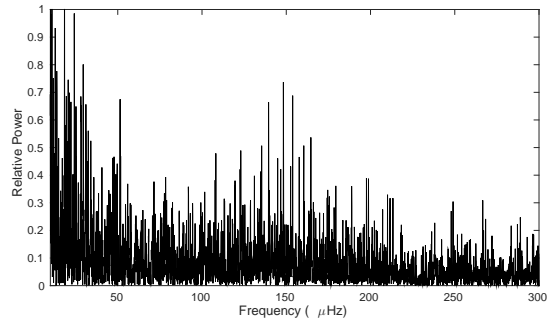} 
 \caption{The power spectrum for the red giant whose time series is shown in Figure~7. Data were processed similarly to those in Stello et al. (2015). A simple quick-look analysis yields $\nu_{max} = 148 \mu Hz$ and $\Delta \nu = 14 \mu Hz$, comparable (within errors) to the results in Stello et al. (2015).}
   \label{fig8}
\end{center}
\end{figure}

\section{Further Directions}

We plan to begin an undergraduate ``student-sourced'' search for transiting exoplanet candidates, concentrating on the secondary sources in the K2 frames, and making use of students taking an ``Exoplanets'' course at Florida Gulf Coast University. In support of this project, we have used Python to construct some simple GUI-based tools which allow visualization and period-searching (using DFT, BLS, etc.) of the K2 time series (Figure~9). The tools run under both Windows and Macintosh operating systems, allowing students to use their personal computers to contribute to the project.

\begin{figure}
\begin{center}
 \includegraphics[width=2.8in]{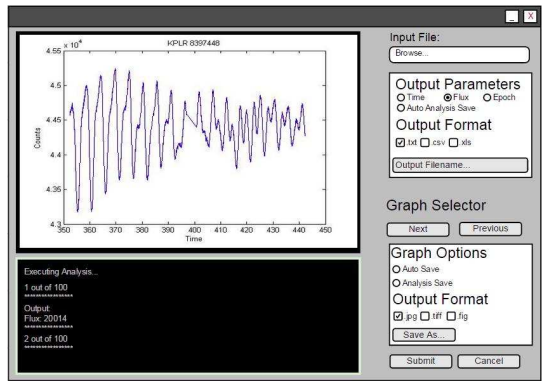} 
 \caption{A screenshot of one of our simplified Python-based tools for visualization and period-finding in the K2 time series. These support undergraduate student exoplanet candidate searches among the K2 secondary targets.}
   \label{fig9}
\end{center}
\end{figure}

In addition, the pipeline supports additional ongoing K2 projects, including rotation/activity correlations among solar analogs, red giant oscillation detection and characterization, and eclipsing binary identification.


\begin{thebibliography}{}

\bibitem[Aigrain \etal\ (2015)]{Aigrain15}
{Aigrain, S. \etal\ }2015,
\textit{MNRAS}, 447, 2880.

\bibitem[Armstrong \etal\ (2015)]{Armstrong15}
{Armstrong, D.J. \etal\ }2015,
\textit{A\&A}, 579, 19.

\bibitem[Box \& Jenkins (1970)]{Box70}
{Box, G. \& Jenkins }1970,
\textit{Time Series Analysis: Forecasting and Control} (San Francisco: Holden-Day).

\bibitem[Gilliland \etal\ (2011)]{Gilliland11}
{Gilliland, R.L. \etal\ }2011,
\textit{ApJS}, 197, 6.

\bibitem[Howell \etal\ (2014)]{Howell14}
{Howell, S.B. \etal\ }2014,
\textit{PASP}, 126, 398.

\bibitem[Lund \etal\ (2015)]{Lund15}
{Lund, M.N. \etal\ }2015,
\textit{ApJ}, 806, 30.

\bibitem[Stello \etal\ (2015)]{Stello15}
{Stello, D. \etal\ }2015,
\textit{ApJL}, 809, 3.

\bibitem[Vanderburg \& Johnson (2014)]{Van14}
{Vanderburg, A. \& Johnson, J.A. }2014,
\textit{PASP}, 126, 948.


\end{thebibliography}
\end{document}